\documentclass{PoS}
\usepackage{amssymb}
\usepackage{booktabs}
\pdfoutput=1

\title{Constraints on polarized parton distributions\\ from open charm and W 
production data}

\ShortTitle{Constraints on polarized PDFs 
from open-charm and $W^\pm$ production data}

\author{\speaker{Emanuele Roberto Nocera}\\
        Universit\`{a}  degli Studi di Milano \& INFN Milano - 
        Via Celoria 16 Milano, Italy\\
        E-mail: \email{emanuele.nocera@unimi.it}}

\abstract{
We study the impact of new open charm muoproduction data from COMPASS
and preliminary $W^\pm$ production data from STAR on {\tt NNPDFpol1.0},
the first unbiased set of polarized parton distributions
recently delivered by the NNPDF Collaboration and 
based on inclusive Deep-Inelastic Scattering data only.
The information contained in the new data sets is incorporated in 
our Monte Carlo parton determination via Bayesian reweighting, a 
method based on statistical inference which avoids the
need for a global refitting. 
We explicitly show to which extent COMPASS and STAR data
can improve our knowledge of gluon and antiquark polarized 
parton distributions respectively.
}

\FullConference{XXI International Workshop on Deep-Inelastic Scattering 
and Related Subjects - DIS2013\\
22-26 April, 2013\\
Marseille, France}

\begin{document}

\paragraph{Introduction.}
In those last years an increasing effort has been devoted to the determination
of helicity-dependent, or polarized, Parton Distribution Functions (PDFs), 
motivated by the role they play in understanding nucleon's spin in terms 
of its Quantum Chromodinamics (QCD) parton substructure~\cite{Leader}.
Spurred by remarkable experimental progress, several next-to-leading order
(NLO) studies were performed, showing which aspects of polarized PDFs 
are probed by the various pieces of available experimental information 
(see for example~\cite{Spinrev} and references therein).
  
The NNPDF Collaboration has recently presented the first unbiased 
determination of polarized PDFs from polarized inclusive Deep-Inelastic 
Scattering (DIS) world data, based on a methodology which uses 
Monte Carlo sampling of experimental data and Neural Networks 
as flexible interpolants~\cite{NNPDFpol}. 
In this analysis, they showed that available inclusive DIS data 
weakly constrain the gluon PDF, probed by scaling
violations in this process, 
unless one makes rather strong assumptions on its 
functional form. Furthermore, presently 
available neutral-current DIS data  
do not allow one to separately determine polarized quark and 
antiquark distributions in the nucleon. 

In this contribution, we study to which extent recent measurements 
of open charm muoproduction data from COMPASS~\cite{CompassOC} 
and preliminary results on $W^\pm$ production data from 
STAR~\cite{STAR} can improve our knowledge 
of the gluon and antiquark polarized PDFs, respectively. 
In order to include this new experimental information in our 
previous fit~\cite{NNPDFpol}, we use 
the reweighting technique~\cite{NNPDFRW1,NNPDFRW2}, which consists of assigning
to each replica in a prior PDF ensemble a weight proportional to the 
probability that this replica agrees with new data. The reweighted ensemble 
is then a representation of the probability distribution of PDFs conditional
on both old and new data. The reweighting method has the main advantage of
avoiding full refitting, which is tipically a cumbersome and time consuming 
task.

\paragraph{COMPASS open charm data.}
The COMPASS Collaboration has recently presented new experimental results 
for the virtual photon-nucleon asymmetry $A_{LL}^{\gamma N \to D^0 X}$ 
obtained from open charm production tagged by $D^0$ meson decays 
in muon-nucleon scattering~\cite{CompassOC}. 
These data potentially provide a direct insight into the polarized 
gluon distribution. At leading-order (LO), open charm production is driven 
by the photon-gluon fusion (PGF) mechanism, $\gamma^*g\to c\bar{c}$, and
the LO virtual photon-nucleon asymmetry $A_{LL}^{\gamma N \to D^0 X}$ 
can be expressed as
\begin{equation}
A_{LL}^{\gamma N \to D^0 X}\equiv\frac{d\Delta\sigma_{\gamma N}}{d\sigma_{\gamma N}}
      =\frac{d\Delta\hat{\sigma}_{\gamma g}\otimes\Delta g\otimes D_c^{D^0}}
            {d\hat{\sigma}_{\gamma g}\otimes g\otimes D_c^{D^0}}
\mbox{ ,}
\label{eq:ALLgammaN}
\end{equation}
where $\Delta\hat{\sigma}_{\gamma g}$ ($\hat{\sigma}_{\gamma g}$) is 
the spin-dependent (-averaged) partonic cross section, 
$\Delta g$ ($g$) is the polarized 
(unpolarized) gluon PDF, 
and $D_{c}^{D^0}$ is the non-perturbative fragmentation
function of a produced charm quark into the observed $D^0$ meson, 
which is assumed spin independent. 
In this framework, COMPASS kinematics allows one to
investigate the polarized gluon PDF 
in the momentum fraction region $0.06 \lesssim x \lesssim 0.22$
and at the energy scale 
$Q^2 = 4(m_c^2+p_T^2) \sim 13$ GeV$^2$, where $m_c$ is the charm quark mass 
and $p_{T}$ is the transverse momentum of the produced charmed hadron.

\begin{figure}[t]
\centering
\includegraphics[scale=0.185]{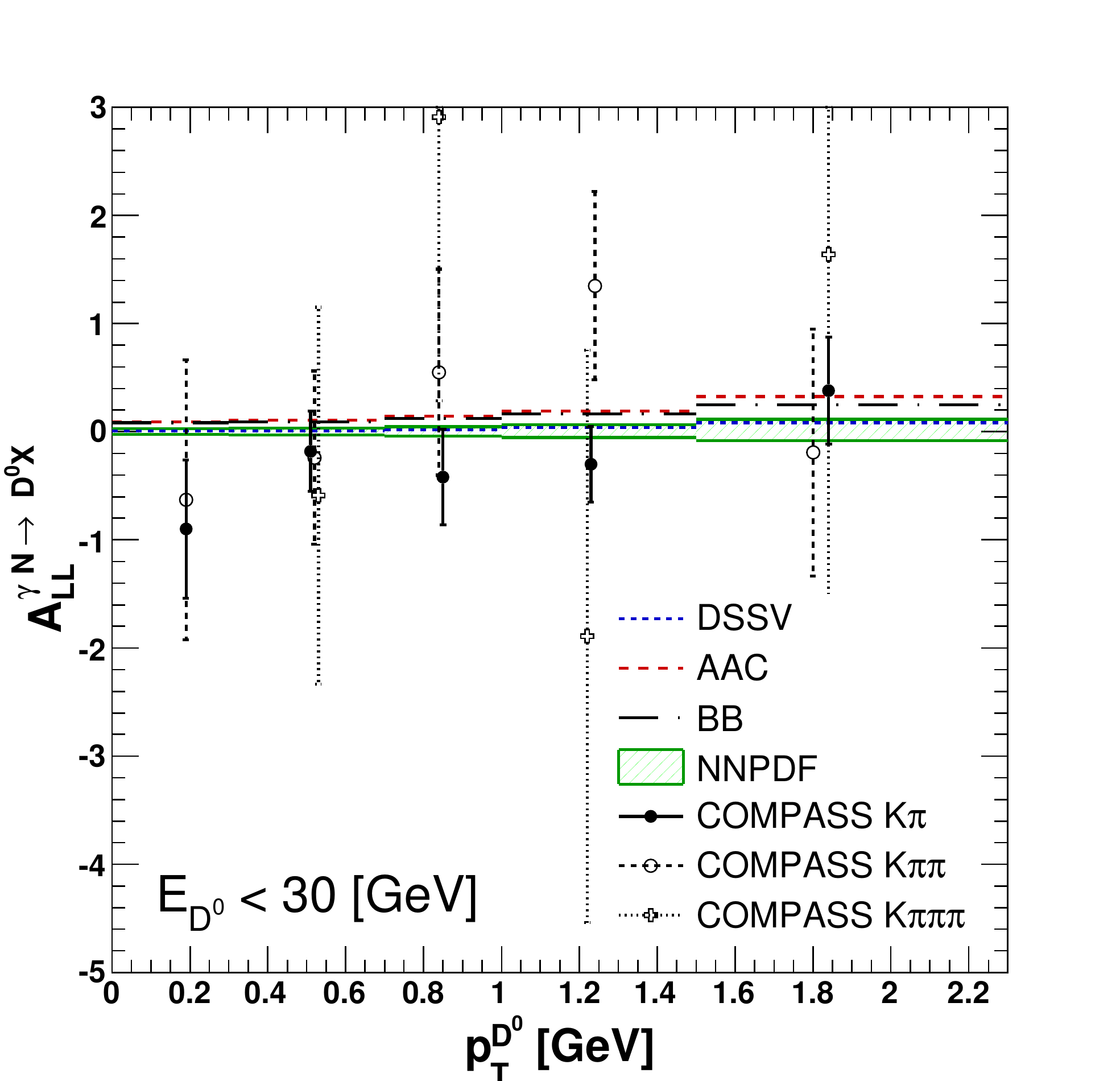}
\includegraphics[scale=0.185]{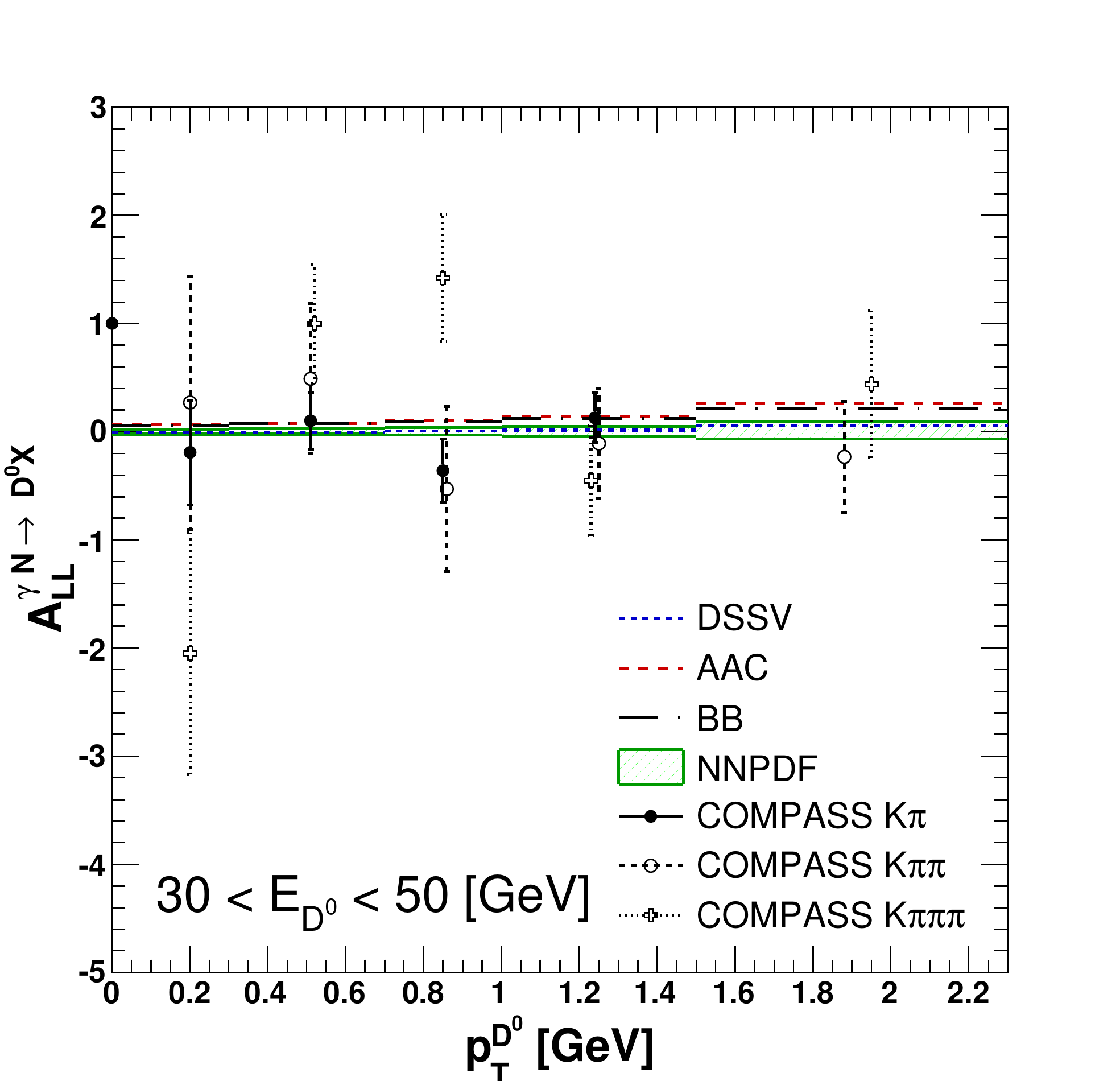}
\includegraphics[scale=0.185]{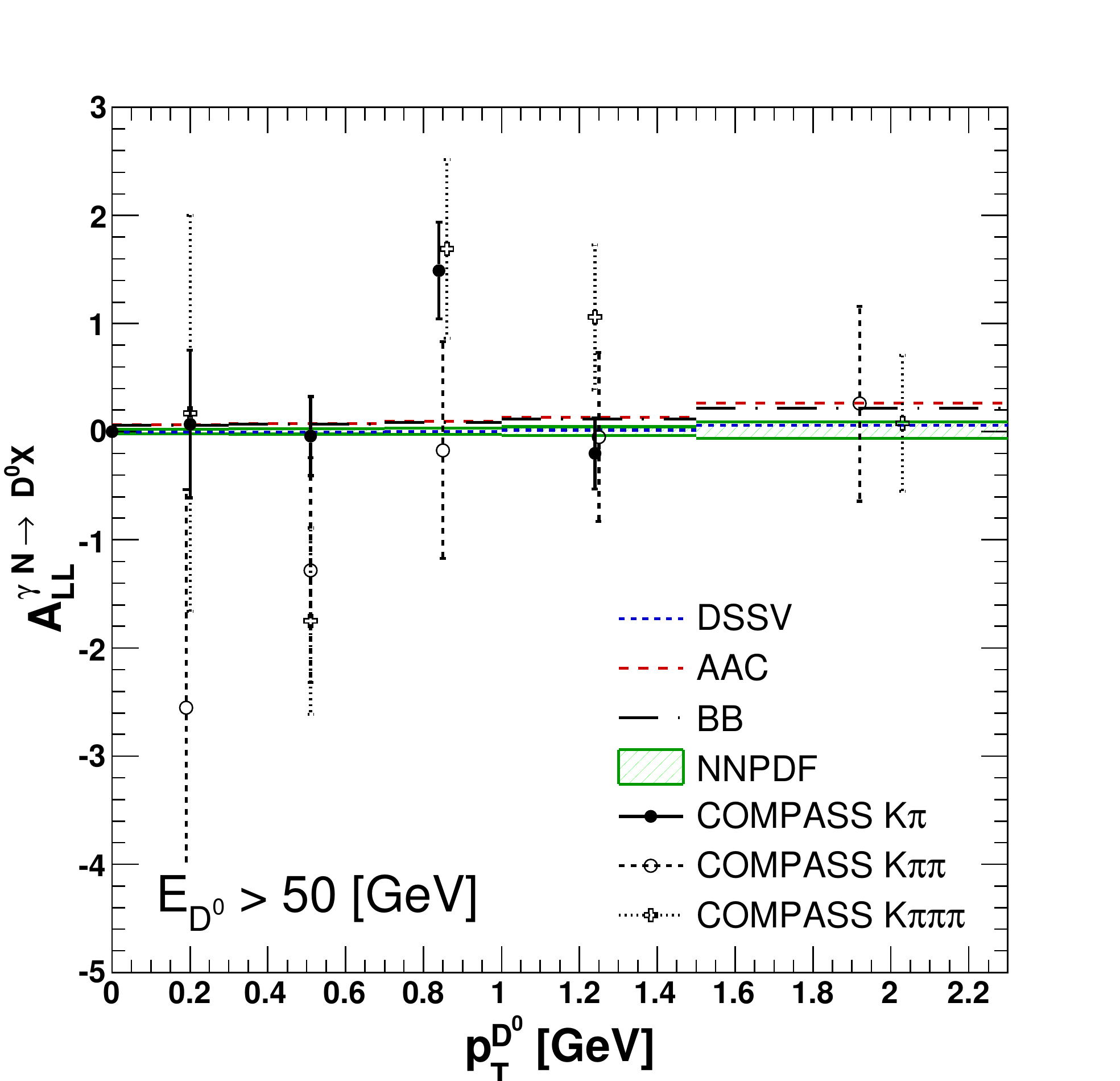}
\caption{Experimental double-spin asymmetry for $D^0$ meson photoproduction
$A_{LL}^{\gamma N \to D^0 X}$ measured by COMPASS~\cite{CompassOC} 
from three decay channels compared to LO predictions computed for different 
PDF sets in three bins of the charmed hadron energy $E_{D^0}$ and in five bins
of its transverse momentum $p_T^{D_0}$.}
\label{fig:CMPOC_obs}
\end{figure}

Theoretical predictions for the virtual photon-nucleon asymmetry 
$A_{LL}^{\gamma N \to D^0 X}$  are made following Ref.~\cite{Simolo} 
by evaluating the analytical expression for both the numerator and the 
denominator in Eq.~(\ref{eq:ALLgammaN}). In Fig.~\ref{fig:CMPOC_obs} we compare 
COMPASS measurements~\cite{CompassOC}, separated into individual decay 
channels, with our predictions for different input parton sets and 
in three bins of the charmed hadron energy; the curve labelled as NNPDF 
is computed using the polarized (unpolarized) gluon
PDF from the \texttt{NNPDFpol1.0}~\cite{NNPDFpol} 
(\texttt{NNPDF2.3 NLO}~\cite{NNPDF23}) parton set. We explicitly checked that 
the asymmetry, both mean value and error, mostly depend on 
the polarized gluon PDF and its uncertainty. 
Predictions do not change if we choose for the unpolarized gluon PDF
a replica at random in the ensemble~\cite{NNPDF23} or
its mean value; also the dependence on the fragmentation
function is moderate, as already pointed out in Ref.~\cite{Riedl}, 
for which we choose the Petersen parametrization as in 
Ref.~\cite{Simolo}.   

\begin{figure}[t]
\centering
\includegraphics[scale=0.185]{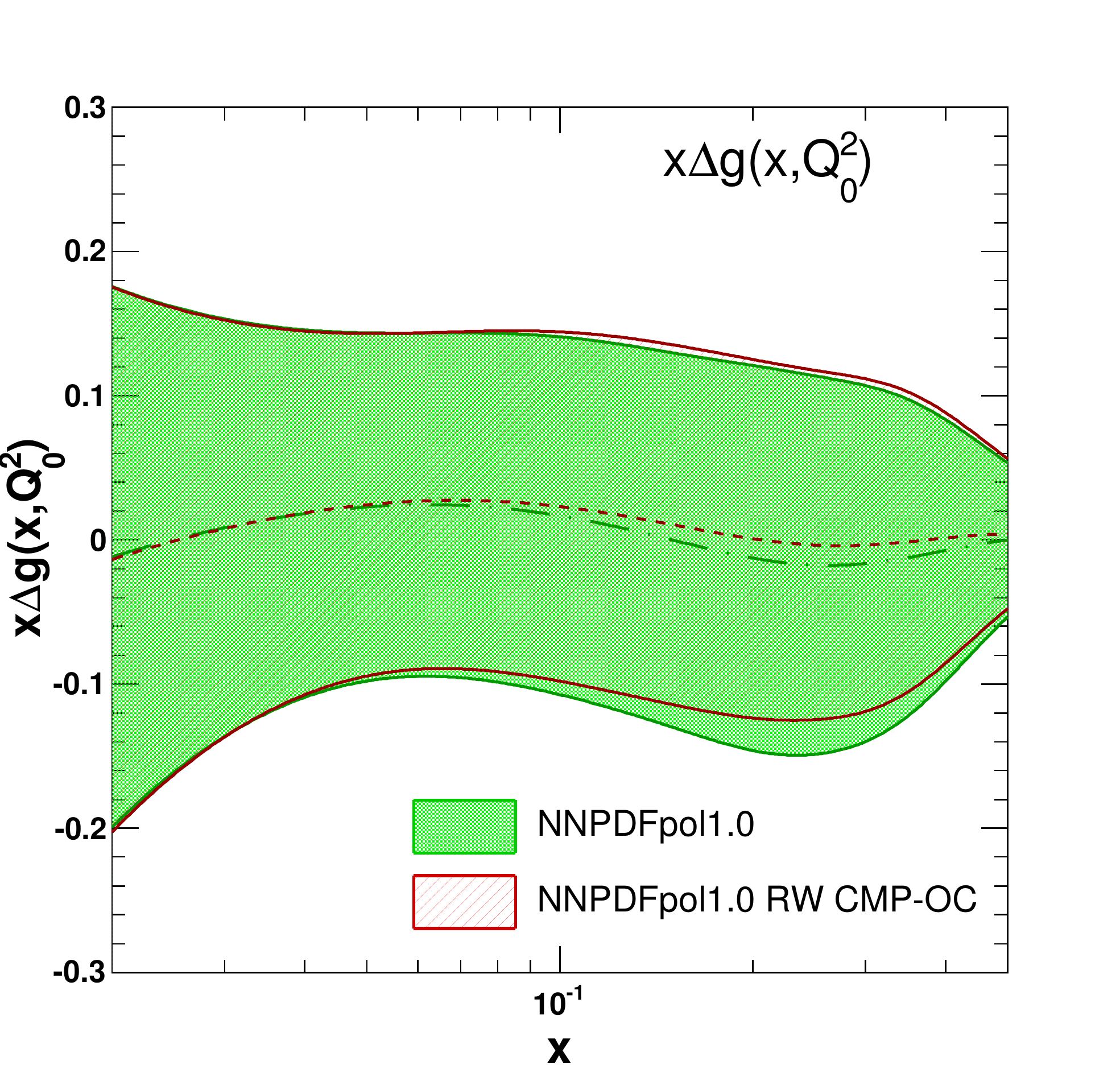}
\includegraphics[scale=0.185]{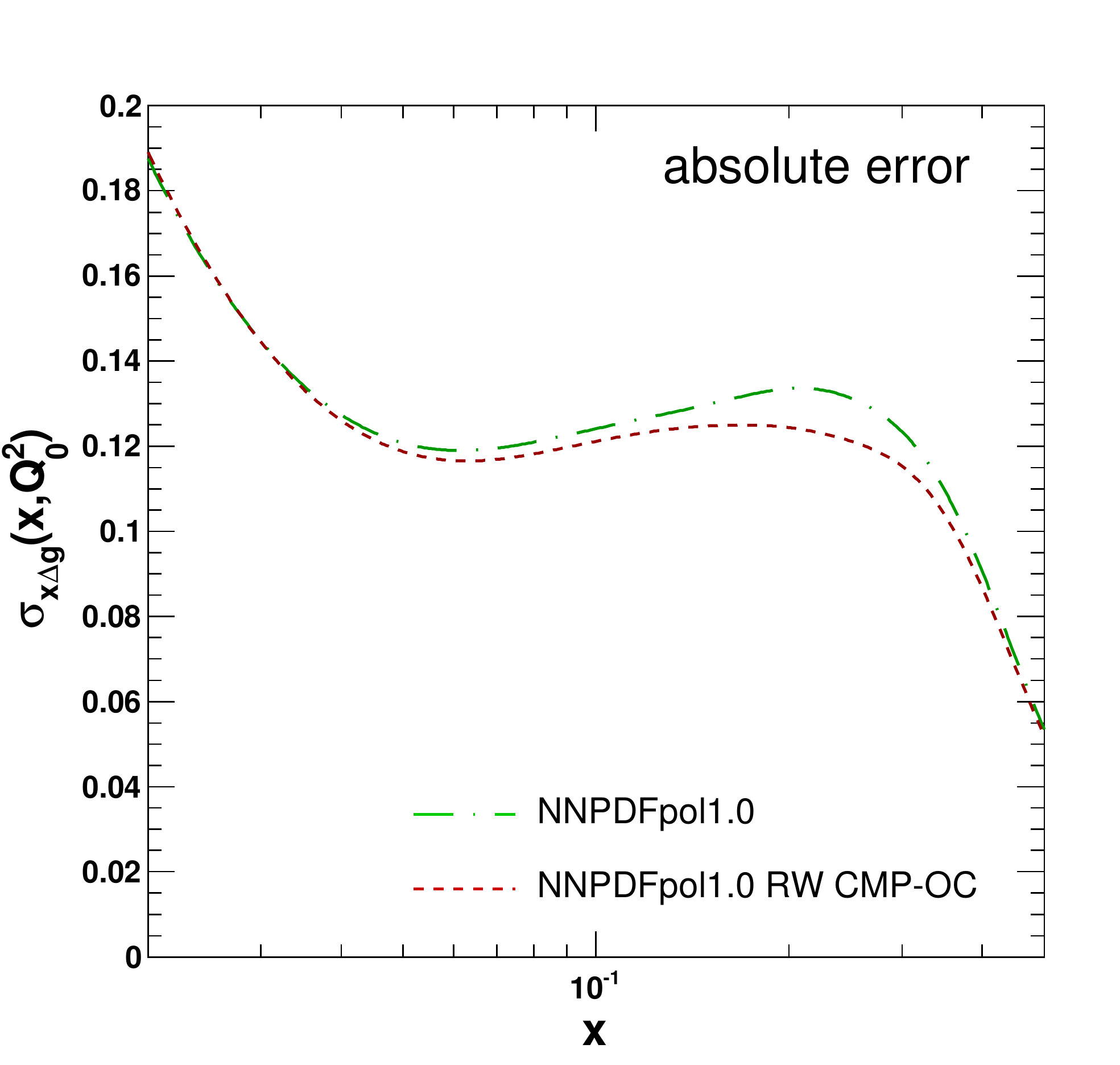}
\caption{Comparison between the unweighted and
the reweighted polarized gluon distribution at $Q_0^2=1$ GeV$^2$ (left panel)
and the improvement in its absolute error (right panel).}
\label{fig:rwgluon}
\end{figure}

We evaluate the $\chi^2$ of the new data to the prediction from each replica 
in the prior PDF ensemble~\cite{NNPDFpol} according to Ref.~\cite{NNPDFRW1}
and obtain the value per data point $\chi^2/N_{\mathrm{dat}}=1.23$, which does 
not decrease after reweighting. We compare the polarized gluon PDF and its 
uncertainty
before and after reweighting in Fig.~\ref{fig:rwgluon}. 
As expected from the present experimental errors
shown in Fig.~\ref{fig:CMPOC_obs}, which are large in comparison to the
theoretical uncertainty due to the PDF, the impact of COMPASS open-charm data
on the polarized gluon PDF is fairly small. The same conclusion was stressed 
in Ref.~\cite{Riedl}, where the longitudinal asymmetry was computed 
at NLO. Even though we computed this asymmetry at LO and its NLO corrections
were shown not to be negligible (see Ref.~\cite{Riedl}), 
we expect our conclusions
will be unchanged when replacing the LO expression, 
Eq.~(\ref{eq:ALLgammaN}), with its NLO counterpart. 
The knowledge on the polarized gluon distribution may be improved by inclusive 
jet production data from the Relativistic Heavy Ion Collider (RHIC), but only
precise measurments at a high-energy Electron-Ion Collider (EIC) might 
reduce its uncertainty in a significant way~\cite{Spinrev}.

\paragraph{STAR W$^\pm$ data.} 
Preliminary measurements on the longitudinal single-spin asymmetry for $W^\pm$
boson production in polarized proton-proton collisions have been presented by 
STAR Collaboration recently~\cite{STAR}. 
Since this asymmetry arises from a purely weak interaction which
couples left-handed quarks with right-handed antiquarks
($u_L\bar{d}_R\to W^+$ and $d_L\bar{u}_R\to W^-$), 
it provides an ideal tool to study individual polarizations 
of quarks and antiquarks inside the proton (see \textit{e.g.} 
Ref.~\cite{Spinrev} for a wider discussion). 
These data are complementary, but independent, 
of those coming from Semi-Inclusive DIS (SIDIS),
however their analysis does not require the usage of poorly
known fragmentation functions.

As already discussed in Ref.~\cite{QCDN12}, the available, DIS-only,
NNPDF polarized parton set~\cite{NNPDFpol} cannot be directly 
used as a prior ensemble for reweighting with STAR-$W$
data sets, since only the sum of quark and antiquark
distributions was fitted there. In Ref.~\cite{QCDN12}, we
argued that suitable prior PDF ensembles may be constructed 
by supplementing the \texttt{NNPDFpol1.0} analysis~\cite{NNPDFpol}
with some reasonable assumption on the antiquark distributions, 
coming for example from existing fits to SIDIS data. If STAR data
bring in sufficient amount of new information, the reweighted PDFs will be
almost independent of the choice of the prior.

Following the method illustrated in Ref.~\cite{QCDN12}, we construct 
four different prior ensembles, each made of $N_{\mathrm{rep}}=300$ 
PDFs, in which the $\Delta\bar{u}$ and 
$\Delta\bar{d}$ distributions are generated from a neural network fit
to the DSSV08~\cite{DSSV08} best fit plus 1-$\sigma$, 2-$\sigma$, 3-$\sigma$,
4-$\sigma$ uncertainty. 
The idea is to enlarge the uncertainty until the reweighted observable 
becomes independent of the prior PDF set: 
we will find that the 3-$\sigma$ and 4-$\sigma$ prior sets 
accommodate our purpose, as discussed below.
Using the CHE code~\cite{CHE}, we then compute for each prior
the electron (positron) longitudinal single-spin asymmetry $A_L^{e^-}$ 
($A_L^{e^-}$) from $W^{-(+)}$ boson production at NLO 
and with STAR kinematics (center-of mass-energy $\sqrt{s}=510$ GeV and
integrated lepton transverse momentum $25<p_T<50$ GeV).

In Tab.~\ref{tab:rwSTAR} we show, for each prior, the 
$\chi^2$ per data point $\chi^2/N_{\mathrm{dat}}$
($\chi^2_{\mathrm{rw}}/N_{\mathrm{dat}}$) before (after) reweighting 
and the fraction $N_{\mathrm{eff}}/N_{\mathrm{rep}}$ of replicas left 
after reweighting, as defined in Ref.~\cite{NNPDFRW1}. 
\begin{table}[t]
  \centering
  \scriptsize
  \begin{tabular}{ll|c|cccc|cccc|cccc}
   \toprule
    Experiment & Set & $N_{\mathrm{dat}}$ 
    & \multicolumn{4}{c|}{$\chi^2/N_{\mathrm{dat}}$}
    & \multicolumn{4}{c|}{$\chi^2_{\mathrm{rw}}/N_{\mathrm{dat}}$}
    & \multicolumn{4}{c}{$N_{\mathrm{eff}}/N_{\mathrm{rep}}$}\\
   \midrule
   & & & 1-$\sigma$ & 2-$\sigma$ & 3-$\sigma$ & 4-$\sigma$ 
       & 1-$\sigma$ & 2-$\sigma$ & 3-$\sigma$ & 4-$\sigma$
       & 1-$\sigma$ & 2-$\sigma$ & 3-$\sigma$ & 4-$\sigma$\\
     STAR     & & 12 & 2.16 & 2.21 & {\bf 2.10} & 2.05 
                     & 1.80 & 1.33 & {\bf 1.06} & 1.06
                     & 0.42 & 0.40 & {\bf 0.38} & 0.32\\
   & STAR $W^+$ &  6 & 1.44 & 1.58 & {\bf 1.58} & 1.57
                     & 1.25 & 1.12 & {\bf 1.04} & 1.04
                     & 0.76 & 0.65 & {\bf 0.60} & 0.50\\
   & STAR $W^-$ &  6 & 2.89 & 2.85 & {\bf 2.62} & 2.53
                     & 2.35 & 1.54 & {\bf 1.08} & 1.07
                     & 0.27 & 0.30 & {\bf 0.30} & 0.27\\
   \bottomrule
   \end{tabular}
\caption{A summary of the reweighting with the $W$ 
data from STAR experiment~\cite{STAR}:
the number of available datapoints $N_{\mathrm{dat}}$, the value of the 
$\chi$ per data point $\chi^2/N_{\mathrm{dat}}$
($\chi^2_{\mathrm{dat}}/N_{\mathrm{dat}}$) before (after) reweighting 
and the fraction $N_{\mathrm{eff}}/N_{\mathrm{rep}}$ of replicas left 
after reweighting ($N_{\mathrm{rep}}=300$), for the different priors  
discussed in the text.}
\label{tab:rwSTAR}
\end{table}
The quality in the description of STAR data
before reweighting is comparable for all prior sets, 
though it is far from being optimal,
since the value of the $\chi^2$ per data point is significantly above one. 
As expected, this value improves after reweighting, 
but different lowering is reached depending on the prior. 
A value of reweighted $\chi^2$ per data point of order one, which
reveals a good description of data, is reached only using 
the 3-$\sigma$ or 4-$\sigma$ prior PDF ensemble.
Notice that no significant improvement in the description of
data is reached moving from the 3-$\sigma$ to the 4-$\sigma$ prior,
thus ensuring the independence of our results from any of these
two choices (and from any other prior constructed assuming less 
stringent antiquark distributions). 
Of course, the latter is less efficient at representing 
the underlying PDF probability distribution, as shown by the lower value 
of the effective number of replicas (see Tab.~\ref{tab:rwSTAR}), 
and a wider PDF ensemble would be needed.

\begin{figure}[t]
\centering
\includegraphics[scale=0.185]{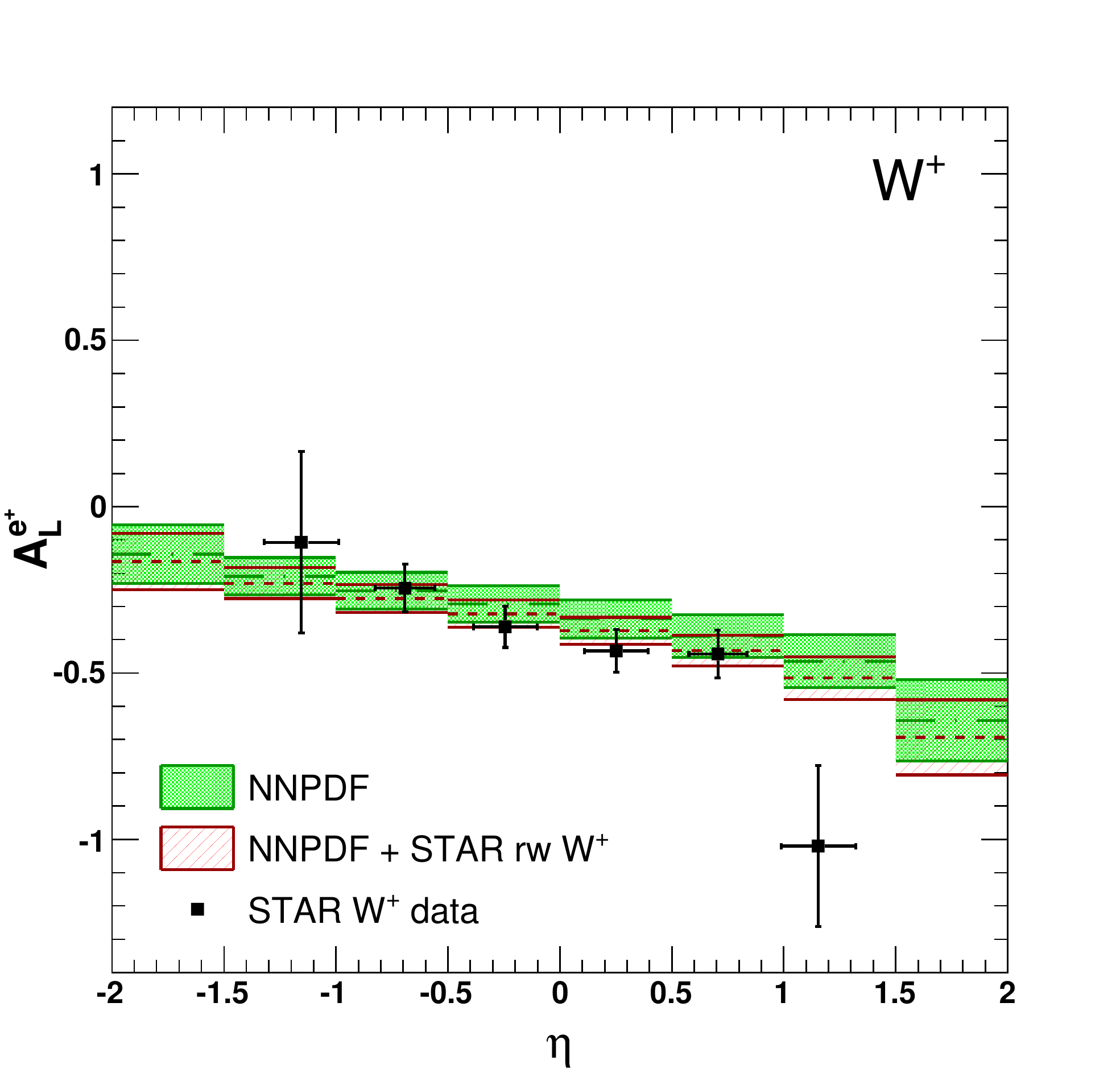}
\includegraphics[scale=0.185]{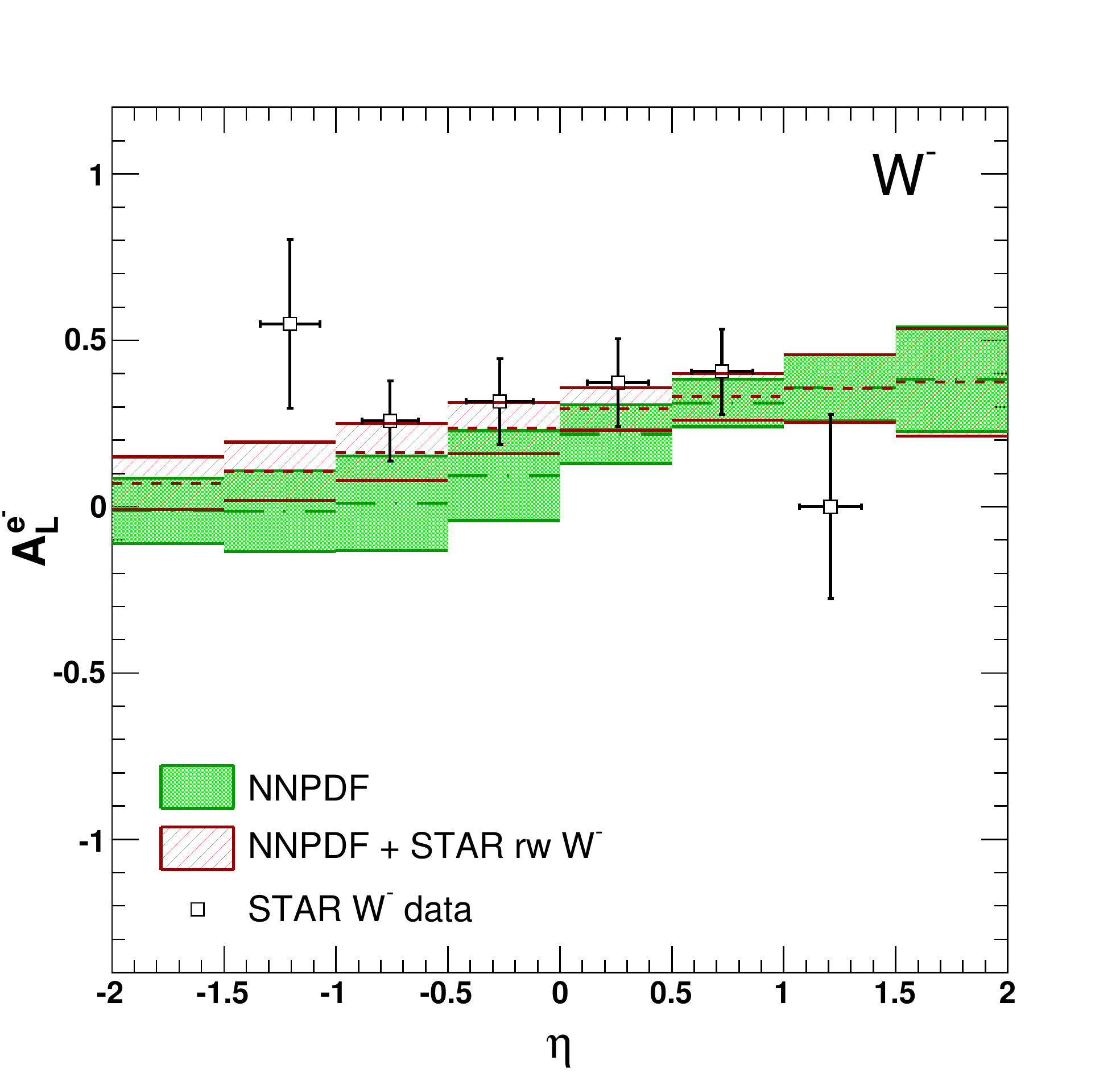}
\includegraphics[scale=0.185]{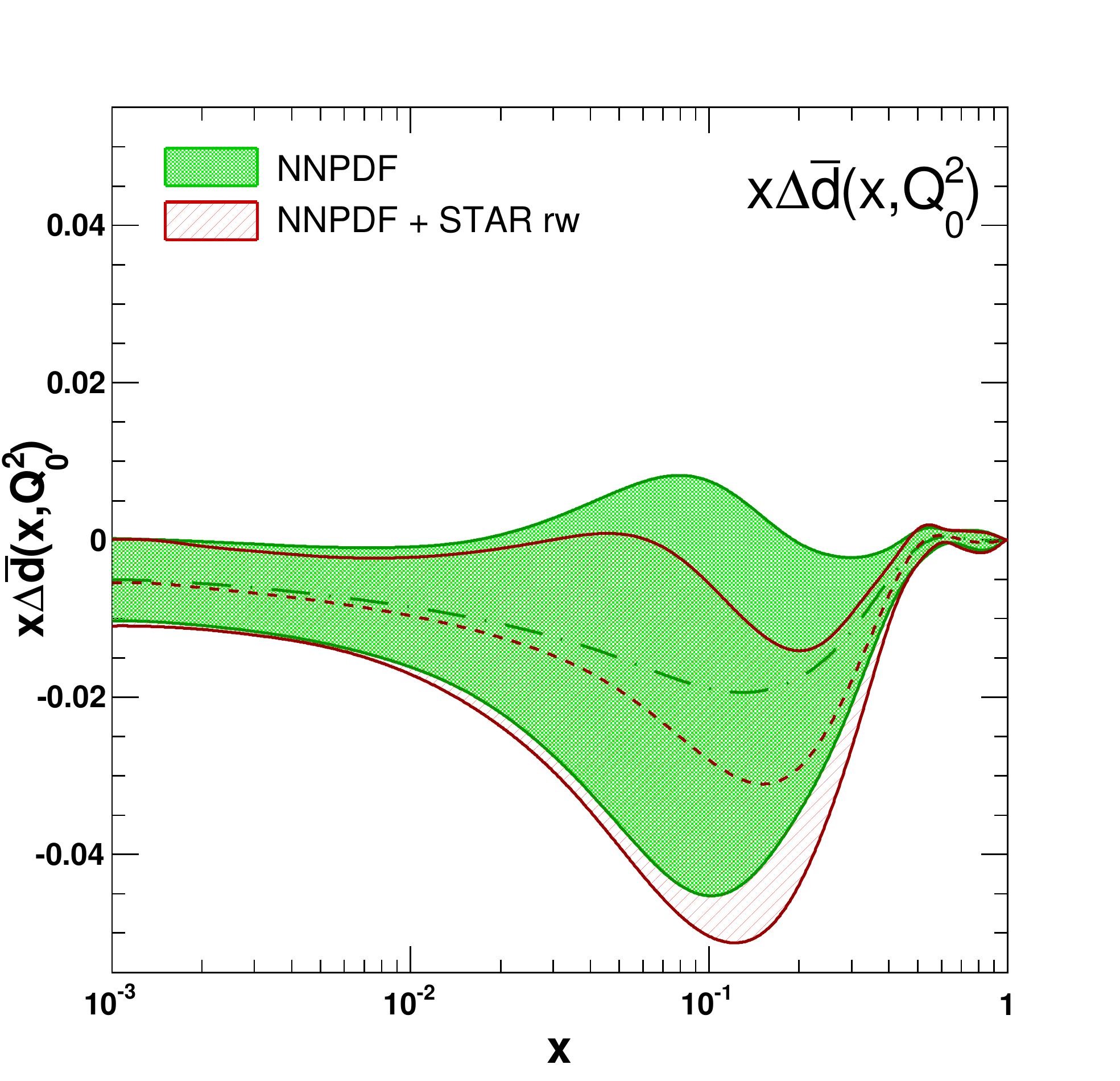}
\includegraphics[scale=0.185]{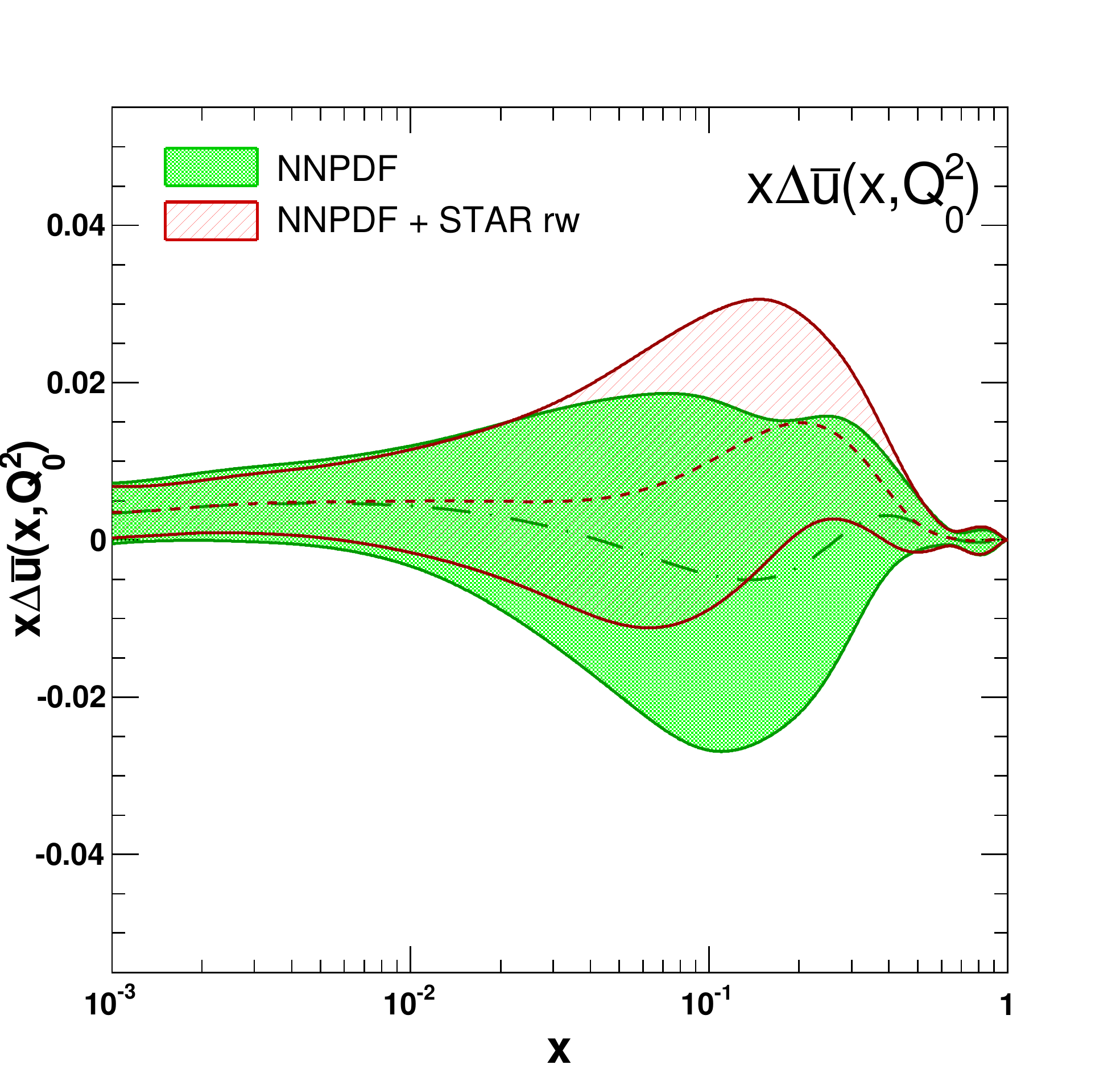}
\caption{The longitudinal positron (electron)
single-spin asymmetry $A_L^{e^+}$ ($A_L^{e^-}$) from
production and decay of $W^{+(-)}$ bosons at STAR
before and after reweighting of 3-$\sigma$ prior, 
in bins of the lepton rapidity $\eta$ and 
toghether with experimental data. The corresponding
probed antiquark PDFs at $Q_0^2=1$ GeV$^2$ are also shown.}
\label{fig:STAR}
\end{figure}

In Fig.~\ref{fig:STAR} we plot the longitudinal positron (electron)
single-spin asymmetry $A_L^{e^+}$ ($A_L^{e^-}$) from
production and decay of $W^{+(-)}$ bosons at STAR. The asymmetry is shown
before and after reweighting of the 3-$\sigma$ prior (boldface in 
Tab.~\ref{tab:rwSTAR}), in bins of the lepton rapidity $\eta$ and 
toghether with experimental data. The reweighted
observable nicely agrees with experimental data and its uncertainty is  
reduced with respect to the prior. In Fig.~\ref{fig:STAR}
we also show the $\Delta\bar{d}$ and the $\Delta\bar{u}$ parton
distributions before and after reweighting at $Q^2=1$ GeV$^2$; 
other PDFs not shown here, including strangeness, are almost unaffected.
All the results shown in Fig.~\ref{fig:STAR} are obtained for the 
simultaneous reweighting with STAR $W^\pm$ data; by reweighting with
separate $W^+$ ($W^-$) data set, we have explicitly checked that 
$\Delta\bar{d}$ ($\Delta\bar{u}$) are separately probed, as expected.

In conclusion, our analysis, though based on a small preliminary
set of $W$ production data, clearly demonstrates their potential 
in constraining the polarized antiquark PDFs; further relevant insight
is expected to be supplied by the 
ongoing analyses and runs from both STAR and PHENIX experiments 
in the forthcoming years.

\acknowledgments
I would like to thank the convenors of the spin working group
for the opportunity to present this work. I am also grateful to
M.~Stratmann for useful discussions and to S.~Forte and J.~Rojo
for a careful reading of the manuscript.

\end{document}